\documentclass{godel-book}

\theoremstyle{definition}

\theoremstyle{remark}

\numberwithin{equation}{section}

\newcommand{\vp}{{\varphi}}
\newcommand{\Vac}{{\Omega}}
\newcommand{\D}{{\partial}}

\newcommand{\al}{{\alpha}}
\newcommand{\la}{{\lambda}}
\newcommand{\w}{{\omega}}
\newcommand{\zb}{{\overline z}}
\newcommand{\Db}{{\overline \D}}

\newcommand{\hh}{{\mathbb H_2}}
\newcommand{\ba}[1]{{\overline{#1}}}

\usepackage{graphicx}

\copyrightinfo{}
{}

\begin{document}

\title{G\"odel-type spacetimes: History and new developments -- Scalar
wave equation in spacetimes of G\"odel type}

\author{Piotr Marecki}
\address{Institute for Theoretical Physics, Leipzig University, Leipzig, Germany}
\curraddr{Institute for Theoretical Physics, Leipzig University, Postfach 100 920, D-04009 Leipzig, Germany}
\email{pmarecki@gmail.com}
\thanks{}

\date{January 1, 1994 and, in revised form, June 22, 1994.}

\keywords{Goedel-type spacetimes, d'Alembert equation}

\begin{abstract}
We analyze the d'Alembert equation in the G\"odel-type spacetimes with spherical and Lobachevsky sections. By separating the $t$ and $x_3$ dependence we reduce the problem to a group-theoretical one. In the spherical case solutions have discrete frequencies, and involve spin-weighted spherical harmonics. In the Lobachevsky case we give simple formulas for obtaining all the solutions belonging to the $D^\pm_\la$ sectors  of the irreducible unitary representations of the reduced Lorentz group. The wave equation enforces restrictions on $\la$ and the allowed (here: continuous) spectrum of frequencies.
\end{abstract}

\maketitle

\section{Introduction}
 In this chapter by G\"odel-type spacetimes we mean homogeneous stationary spacetimes in which dust\footnote{Dust is one of the components of possible sources, $T_{ab}$, of G\"odel spacetimes; see \cite{RG}.} rotates around every point relative to the ``compass of inertia'' \cite{Goedel,RG}. As in isotropic cosmology there are three qualitatively different types of G\"odel spacetimes, constructed upon homogeneous three-dimensional spaces with either:  Lobachevsky ($\mathbb H_2\times \mathbb R$),  spherical ($\mathbb S_2\times \mathbb R$) or flat ($\mathbb R^3$) geometry. These spaces will be called here  \emph{homogeneous sections}. 

Formally, all these spacetimes possess the parametrization
\begin{equation}\label{general_form}
 d(Rs)^2=(dt+A_i dx^i)^2 -h_{ij} dx^i dx^j,
\end{equation}
where $i=1,\ldots, 3$ and the ($t-$independent) vector field $A^i$ possesses a constant rotation $\Omega$ on the homogeneous sections, the metric of which is denoted by $h_{ij}$. For a chosen homogeneous section the scale $R$ and the rotation $\Omega$ are the only parameters specifying the spacetime under consideration.

In this chapter we will review the construction of solutions of the scalar d'Alem\-bert equation in G\"odel-type spacetimes. It is intuitively clear, that the task is similar to the Landau problem in quantum mechanics, i.e. to the Schr\"odinger problem in a static, homogeneous magnetic field. Indeed the analogy is almost complete: as in quantum mechanics there are two convenient ansaetze -- exhibiting either the plane-wave structure in one of the dimensions or the vortex-like structure of the solutions. Here we shall follow the second one. Moreover, we are of the opinion that algebraic solutions are the most elegant ones and we shall exhibit a solution of this type here. By ``algebraic'' we mean the construction of solutions based on the structure of the underlying symmetry-group, which builds the whole family of solutions, belonging to specified sectors of irreducible representations of the group, with the help of ladder-operators.

\subsection{Approaches to the problem}

In the literature Drukker, Fiol and Sim\'on \cite{DFS} seem to have gone the furthest to employ the algebraic approach\footnote{The group structure is implicit in their approach; note, however, incorrect characterization of the solutions in the spherical case, where the underlying group $SO(3)$ is compact and the spectrum of the Casimir operator $L^2$ necessarily discrete.}. Also relevant is the work of Figueiredo, Damiao-Soares and Tiomno \cite{FST} who have constructed the solutions algebraically, but have left them in the form of hypergeometric functions, while a simple description in terms of complex variables exists (also noted by \cite{DFS}) and will be presented here. In \cite{WH} the d'Alembert equation for the original G\"odel spacetime (Lobachevsky-type) was solved in the planar approach\footnote{Note, however, the sign error in front of $\w^2$ in this paper.}. The papers of Dunne \cite{GD} and Comtet \cite{AC} treat algebraically the Schr\"odinger pro\-blem for a particle in a homogeneous magnetic field on a homogeneous space, and are relevant to the present problem as well.

\section{Solving the wave equation by an ansatz}

In order to find the solutions of the d'Alembert equation in the spacetimes of G\"odel type we will make use of the fact that these spacetimes are stationary, homogeneous and axially symmetric. The three cases mentioned in the introduction are obtained by setting
\begin{equation}
d(Rs)^2=\bigl(dt+H(x) d\vp\bigr)^2-dx^2-D^2(x)d\vp^2-(dx_3)^2 
\end{equation}
with
\begin{equation}
H(x)=\begin{cases}2\al \sinh^2(x/2), \\ 2\al \sin^2(x/2), \\ \frac{1}{2}\,\al x^2, 
\end{cases} \qquad
D(x)=\begin{cases}\sinh(x), \\ \sin(x), \\  x, 
\end{cases}
\end{equation}
for Lobachevsky, spherical and flat homogeneous sections respectively\footnote{The meaning of constants is discussed below.}.
In solving the d'Alembert equation,
\begin{equation}
\Box \Psi (t,r,\vp,x_3)=0
\end{equation}
we shall employ the separation ansatz reflecting the translational symmetry in $t$ and $x_3$ directions:
\begin{equation}\label{separation}
\Psi(t,x,\vp,x_3)=e^{-i\w t}e^{ik_3x_3}\cdot \psi(x,\vp),
\end{equation}
where the function $\psi(x,\vp)$ should be single-valued (periodic) function of $\vp$ with the period of $2\pi$. The axial symmetry simplifies the problem even further to a single second order ODE (in $x$), subject to the following consideration of boundary conditions. For smooth functions of compact support the d'Alembert operator, by standard arguments, is symmetric\footnote{We are not aware of any rigorous functional-analytic study of the selfadjointness of this operator on G\"odel-type spacetimes.} with respect to the scalar product
\begin{equation}
(F,G)=\int \sqrt{-g}\, d^4 x\,\, \ba{F}  G=\int dt\, dx_3\, D(x) dx d\vp\,\, \ba{F} G. 
\end{equation}
On the other hand, considering functions on homogeneous two-dimensional Loba\-chev\-sky $\mathbb H_2$, spherical $\mathbb S_2$ and flat $\mathbb R_2$ spaces, one has the standard scalar product $(.,.)_2$ of these spaces:
\begin{equation}
(f,g)_2=\int D(x)dx d\vp\,\, \ba{f} g.
\end{equation}
This is all possible due to $-\det (g_{ab})=\det (h_{ij})\equiv h$. Below we shall use the product $(.,.)_2$ to derive the boundary conditions (square-integrability) supplementing the second order ODE discussed above. The algebraic method advertised in the introduction, which determines the $\psi(x,\vp)$ without solving the second-order ODE, uses a deformation of symmetry generators of $\hh$ and $\mathbb S_2$. These operators are selfadjoint on the Hilbert spaces of square-integrable functions  with the scalar product $(.,.)$, as described in appendix A.

Let us now discuss the Lobachevsky and spherical cases separately. 

\subsection{Solution in Lobachevsky case}\label{lob}
Let us consider the spacetimes with Lobachevsky homogeneous sections first, as the original G\"odel spacetime \cite{Goedel} is of this type.  These spacetimes are parameterized, using the standard notation \cite{FST,DFS}, by the ``amount of rotation'' $\Vac$ and the magnitude of Lobachevsky curvature denoted by $l$ :
\begin{equation}
d(2ls)^2=(dt+H d\vp)^2-dx^2-D^2 d\vp^2-dx_3^2,
\end{equation}
where all the coordinates are dimensionless, and the parameter $\al={\Vac}/{l}$. Both $t$ and $x_3$ range over the whole $\mathbb R$, while $x\in[0,\infty)$ and $\vp\in[0,2\pi)$ with the usual identification of $0$ and $2 \pi$. The metric is of the form \eqref{general_form} with  metric of the homogeneous section being 
\begin{equation}
h_{ij}dx^idx^j=dx^2 + \sinh^2(x)\, d\vp^2+dx_3^2.
\end{equation}
The rotation of the vector field $A_i$ (interpreted as a vector field on the homogeneous section) is a vector field directed along the $x_3$ direction: 
\begin{equation}
B^i=\frac{\epsilon^{ijk}\D_jA_k}{\sqrt{h}}=\al\, \delta^i_3,
\end{equation}
and per analogy with other solutions of Einstein's field equations can be thought of as representing and a constant gravimagnetic (Lense-Thirring) field on this space. The constant $\al$ is also equal to the contravariant $x
_3$ component of the rotation vector of the $(\D_t)^a$ vector field (velocity field of the sources). The case $\Vac=0$ corresponds to a static spacetime with Lobachevsky $(x,\vp)$ sections, while $\Vac^2=2l^2$ (that is $\al=\sqrt{2}$) corresponds to the original G\"odel metric.

For technical reasons we shall derive solutions of the wave equation only in the case of sufficiently strong rotation $\Vac>l$ (that is, $\al>1$). We interpret the solutions  as representing the ``bound states''. In a general case of arbitrary $\Vac$ our solutions are supplemented by a family of ``scattering states'', i.e. solutions of the type to be expected if $\Vac=0$ (see discussion after Eq. \eqref{ode}). 

 In order to solve the wave equation in these spacetimes, we will make use of the underlying Lobachevsky structure. The d'Alembert equation for $\Psi$ is
\begin{eqnarray}
 \Box \Psi&=\left[1-\al^2\tanh^2(x/2)\right] \Psi_{tt}+\frac{\al}{\cosh^2(x/2)}\Psi_{t\vp}-\nonumber \\
&-\frac{1}{\sinh^2(x)}\Psi_{\vp\vp}-\Psi_{33}-\Psi_{xx}-\coth(x)\Psi_x=0.
\end{eqnarray}
The following three operators, which are Killing vectors of the G\"odel-Lobachevsky spacetimes multiplied by $-i$, fulfill the same commutation relations as the Killing vectors Lobachevsky plane also multiplied by $-i$ (cf. the $L_i$'s in appendix A):
\begin{eqnarray}
J_1&=&{-i}\sin\vp\D_x-i\coth(x)\cos(\vp)\D_\vp-i{\al}\cos\vp\tanh(x/2)\D_t,\\
J_2&=&{i}\cos\vp\D_x-i\coth(x)\sin(\vp)\D_\vp-i{\al}\sin\vp\tanh(x/2)\D_t,\\
J_0&=&-i\D_\vp-i{\al\D_t},
\end{eqnarray}
while the remaining two Killing vectors, $\D_t$ and $\D_3$, obviously commute with the above ones.
Furthermore, we notice that making use of the separation ansatz (\ref{separation}) for $\Psi$, we are lead to
\begin{equation}
\Box \Psi=-C_2\Psi-(1-\al^2)\w^2\Psi-\Psi_{33},
\end{equation}
where here: $C_2=J_0^2-J_1^2-J_2^2$. The wave equation therefore reduces remarkably to
\begin{equation}\label{cdwa}
C_2\psi(x,\vp) =\left[(\al^2-1)\w^2+k_3^2\right]\psi(x,\vp)
\end{equation}
Because of our assumption\footnote{Curiously,  $\al^2>1$ is also the condition for the appearance of closed time-like curves in the spacetime.}  $\al^2>1$, it remains to determine the possible eigenfunctions, $\psi(x,\vp)$, with a \emph{positive} eigenvalue of the Casimir operator $C_2$. The operator $C_2$ is now a second order partial differential operator acting on $\psi(x,\vp)$.

The problem of determining the eigenfunctions can be approach in two ways. The generic way is to use the axial symmetry (and periodicity in $\vp$) and solve the remaining second-order ODE in $x$. Here we shall follow a different, algebraic method, which is simpler. Let us  regard $C_2$ (and, in fact, the whole set $J_a$) as an operator acting in the Hilbert space of square-integrable functions of $(x,\vp)$ with the scalar product provided by the Lobachevsky product, (\ref{product}).

It is helpful to recall the construction of the unitary irreducible representations of the reduced Lorentz group $\mathfrak L_3$ (two boosts and one rotation), which is the symmetry group of the Lobachevsky space $\hh$, and of its corresponding spinor group $\mathfrak S_3\equiv$SU(1,1) of Bargmann \cite{Ba}. For positive eigenvalues of $C_2$ one is lead to what Bargmann calls \emph{discrete classes\footnote{Note, that with our operator $C_2$ corresponds to $-\mathcal Q$ of Bargmann \cite{Ba}. It is worth noting that for positive $C_2$ only the discrete class appears. For negative $C_2$ discrete and continuous classes appear.}}, $D^-_\la$ and $D^+_\la$. Within $D^\pm_\la$ the eigenvalue of $C_2$ is $\la(\la\mp 1)$ and the eigenvalues of $J_0$ start with $\la$ and increase (without bound) by one with each application of $J_+=J_1+iJ_2$ in the $D^+_\la$ class (or, respectively, start with $-\la$ and decrease (without bound) by one with each application of $J_-$, for the $D^-_\la$ class). In $D^+_\la$ the number $\la$ is real and positive\footnote{Bargmann notes (see \S4 of \cite{Ba}), that the universal covering group $\mathfrak C$ covers $\mathfrak S_3$ infinitely often. He restricts himself to single-valued representations of $\mathfrak S_3$, and as a consequence treats only integer and half-integer $\lambda$'s. We see no reason for such a restriction here, and therefore leave $\lambda\in \mathbb R_+$. This has the profound consequence, that the eigenvalues of $C_2$ are no longer discrete in the \emph{discrete class} (this discreteness was the reason for the label ``discrete'' of the discrete class in \cite{Ba}).}, while in $D^-_\la$ it is real and negative. More precisely: $|\la|\geq 1$ in both classes.

Let us determine all these functions explicitly. Introducing the complex independent variable 
\begin{equation}
z=\tanh(x/2)\, e^{i\vp},
\end{equation}
we note that \[\D_\vp=iz\D -i\zb\Db, \] where $\Db=\D_{\zb}$, and  find
\begin{eqnarray}
J_+&=z^2 \D -  \Db +a z ,\\
J_-&= \D - \zb^2 \Db +a \zb,\\
J_0&=z\D-\zb\Db+a,
\end{eqnarray}
where
\begin{equation}\label{param}
a=-\al \w
\end{equation}
and $J_\pm=J_1\pm i J_2$. We may verify, that
\begin{eqnarray}
[J_+,J_-]&=&-2J_0,\\
C_2&=&J_0(J_0+1)-J_-J_+,\\
C_2&=&J_0(J_0-1)-J_+J_-,\label{eigenc2}
\end{eqnarray}
which are all standard relations in the theory of representations of SU(1,1), and thus are the same as the commutation rules among the generators $L_i$ of the symmetry group of $\hh$.

For a given $\la$ the families of simultaneous eigenfunctions of $C_2$ and $J_0$ are constructed upon a reference vectors\footnote{Denoted by $g$ in \S5g of \cite{Ba}.}, denoted here by $\psi^+_0$ (or $\psi^-_0$), annihilated by\footnote{The necessity for such a vector is clear; as in the
case of the rotation group - its absence would lead to
negative-norm states created by successive application of either
$J_-$ (or $J_+$).} $J_-$ (or $J_+$). This leads to two families, $D^+_\la$ and $D^-_\la$.

\subsubsection{Structure of $D^+_\la$.}
We find immediately
\begin{eqnarray}
\psi^+_0&=(1-z\zb)^\la \zb^{a-\la},
\end{eqnarray}
where $\la$ is the eigenvalue of $J_0$, and $\la(\la-1)$ the eigenvalue of $C_2$. The function $\psi_0^+$ depends, via the last term, on the angle $\vp$. Since it must be periodic in $\vp$, with identical values for $\vp=0$ and $\vp=2\pi$, the number $a-\la\equiv m$ must be an integer. In the original variables we have
\begin{equation}
\psi^+_0=\cosh(x/2)^{-2\la-m}\sinh(x/2)^m\, e^{-im\vp}.
\end{equation}
Further restrictions,
\begin{equation}
m\geq 0, \qquad \la>\frac{1}{2},
\end{equation}
are obtained from the requirement of square integrability w.r.t. the product (\ref{product}) at $x=0$ and at infinity, respectively\footnote{Solutions not fulfilling this condition are growing exponentially for large $x$.}.

A connection with the methods of ordinary differential equations is provided by the following consideration: the eigenvalue problem for the Casimir operator,
\begin{equation}
C_2\psi=\la(\la-1)\psi,
\end{equation}
with $\la\geq 1$ leads, with a separation ansatz $\psi=f(x)\exp(-im\vp)$, $\mathbb Z\ni m\geq 0$ to the equation
\begin{equation}\label{ode}
w(w-2)f'' +2(w-1)f'+\left[\frac{2a(a+m)}{w}-\frac{m^2}{w(w-2)}\right]f=\la(\la-1)f,
\end{equation}
where $w=2\cosh^2(x/2)$. This is an equation with three regular singular points, at $(0,2,\infty)$, with the exponents $(a+\frac{m}{2},a-\frac{m}{2})$, $(\frac{m}{2},-\frac{m}{2})$ and $(\la-1,-\la)$ respectively. Evidently, the function(s) we have found by algebraic methods, $\psi^+_0$, have the right asymptotics at $w=2$ ($\psi^+_0\sim (w-2)^\frac{m}{2}$) \emph{and} at infinity ($\psi^+_0\sim w^{-\la}$) to be square integrable (w.r.t. \eqref{product}) at these extremes.

The equation \eqref{ode} is, as expected, also valid in the case of slower rotation $\Vac\leq 1$, i.e. $\al<1$. Thus, in this case, the solutions are of the form of hypergeometric functions (of $w=2\cos^2(x/2)$) multiplied with $e^{im\vp}$. We see no simpler expressions\footnote{These functions  correspond to eigenfunctions with a negative eigenvalue of $C_2$ of Bargmann. While again one can generate new eigenfunctions by an application of $J_+$ or $J_-$ in this case there does not exist any reference vectors annihilated by $J_+$ or $J_-$, and consequently none of the functions can be determined from first-order differential equations as in the case of $C_2>0$.} for them in terms of the complex variable $z=\tanh(x/2)\, e^{i\vp}$. \\

Let us now finish the construction of solutions in the $D^+_\la$ case. The solutions are characterized by the momentum in the $x_3$ direction ($k_3$), by the positive integer $m$ and by the representation parameters $(\la,k)$ with $k\in\{\la,\la+1,\ldots,\infty\}$. For $k=\la$ the solution coincides with the extremal function $\psi_0^+$ given above. Denoting the solutions with arbitrary numbers $k,\la,m$ by ${_mL^\la_k}(x,\vp)$ we have
\[
 {_mL^\la_\la}(x,\vp)=(1-z\zb)^\la \zb^m.
\]
Solutions with  $k\geq \la$ are found by  successive applications of the operator\footnote{Normalization of these solutions can be achieved straightforwardly.} $J_+$ to ${_mL^\la_\la}$. The frequency of the solutions, here negative, is given by
\[
 \w=-\frac{a}{\al}=-\frac{\la+m}{\al},
\]
and is determined by the parameters $\la,m$ alone. The only role of the wave equation, which reduces to
\begin{equation}\label{def_k}
\la(\la-1)=(1-\tfrac{1}{\al^2}) (\la+m)^2+ k_3^2,
\end{equation}
is to fix the momentum in the $x_3$ direction, i.e. the parameter $k_3$, again as a function of $\la,m$ only.

This equation for each $m$ leads to a restriction on the allowed values of $\la$ so as to make the expression for $k_3^2$ positive. In particular there exists a minimal $\la$,  $\la_{min}=\al^2$; this restriction together with the corresponding one in $D^-_\la$ leads to a gap in frequencies: $|\w|\geq \al$.\\

\subsubsection{Structure of $D^-_\la$.}
The reference vecotr, annihilated by $J_+$, is
\begin{equation}
\psi_0^-=(1-z\zb)^{-\la}\, z^{m},
\end{equation}
with $m=\la-a$. It is an eigenfunction of $C_2$ to the eigenvalue $\la(\la+1)$, and of $J_0$ to the eigenvalue $\la$ (with $\la\leq -1$ in this family). The function $\psi_0^-$ is square-integrable if $\la<-\tfrac{1}{2}$ and if $m\in \mathbb N_0$. The frequencies are positive in this family,
\[
 \w=-\frac{a}{\al}=\frac{-\la+m}{\al},
\]
and $\la,m$ determine the $x_3$-momentum via,
\begin{equation}
\la(\la+1)=(1-\tfrac{1}{\al^2}) (-\la+m)^2+ k_3^2,
\end{equation}
which again restricts $\la$ for a given $m$.

We note, that by a complex conjugation of the (negative-frequency) solution of the wave equation
\[
 e^{-i\w t}e^{ik_3x_3} (_mL^\la_k)(x,\vp),
\]
with $(_mL^\la_k)$ from $D^+_\la$ a positive frequency solution
\[
 e^{-i\w' t}e^{ik_3x_3} (_{m'}L^{\la'}_{k'})(x,\vp),
\]
with $(_{m'}L^{\la'}_{k'})\in D^-_{\la'}$,
\[
 \la'=-\la, \qquad k'=-k, \qquad m'=m,\qquad \w'=-\w
\]
is obtained.

\subsection*{Summary of Lobachevsky case}
The solutions of the d'Alembert equation have the form:
\[
\Psi(t,x,\vp,x_3)=e^{-i\w t}e^{ik_3x_3}\cdot (_mL^\la_k)(x,\vp),
\]
with the following restrictions/specifications on the parameters of the functions $(_mL^\la_k)$:
\begin{table}[h]
\caption{Summary of the Lobachevsky case.}\label{eqtable}
\renewcommand\arraystretch{1.5}
\noindent\[
\begin{array}{|l|c|c|}
\hline
&D^+_\la\ (negative\, frequency)&D^-_\la\ (positive\, frequency) \\
\hline
sector, {\la}&\mathbb R_+\ni \la \geq 1& \mathbb R_-\ni \la \leq -1\\
\hline
eigenvalue \, of\, J_0, \, {k}&k=\la,\la+1,\ldots & k=\la, \la-1,\ldots \\
\hline
eigenvalue \, of \ C_2&\la(\la-1)\geq 0&\la(\la+1)\geq 0\\
\hline
extremal \, vector, \, (_m L^\la_\la) & (1-z\zb)^\la \zb^m& (1-z\zb)^{-\la} z^m\\
\hline
 & \multicolumn{2}{|c|}{ z=\tanh(x/2)e^{i\vp}}
\\
\hline
vectors\ ({_mL^\la_k}) \ generated \ by & J_+ & J_-\\
\hline
parameter\ m&m\in \mathbb N_0& m\in \mathbb N_0\\
\hline
parameter \ {a}& a=\la+m\geq0& a=\la-m\leq 0\\
\hline
frequency \ \w& \w=-\frac{\la+m}{\al}\leq -\al & \w=-\frac{\la-m}{\al}\geq \al\\
\hline
minimal/maximal \ {\la}& \la_{min}=\al^2&\la_{max}=-\al^2\\
\hline
\end{array}
\]
\end{table}

In both cases the values of $\la$ and $m$ specify completely the frequency, and also, by the wave equation, the momentum in the $x_3$-direction:
\begin{align}
k_3&=\pm\sqrt{\la(\la-1)-(1-1/\al^2)(\la+m)^2}, \qquad \text{for}\, D^+_\la\\
k_3&=\pm\sqrt{\la(\la+1)-(1-1/\al^2)(\la-m)^2}, \qquad \text{for}\, D^-_\la
\end{align}
These equations for each $m$ lead restrictions on the allowed values of $\la$, which in turn imply the gap in frequencies: $|\w|\geq \al$.
\begin{figure}[htb]
\includegraphics[scale=0.6]{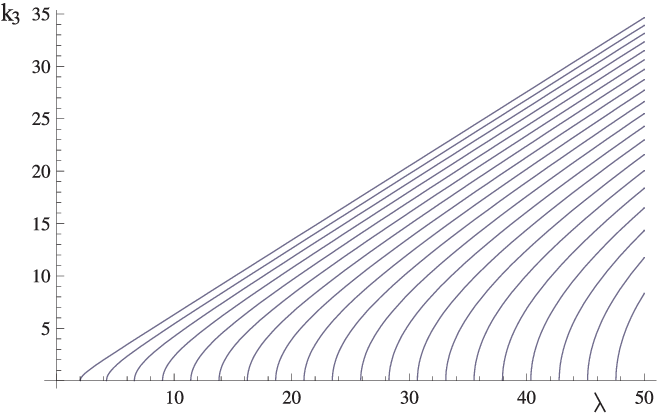}
\caption{Momenta in $x_3$ direction in Lobachevsky $D^+$ case as a function of $\la$ for successive values of $m$.}
\label{firstfig}
\end{figure}\\

\subsection{Solution in spherical case}\label{sph}
In the case of spacetimes with  spherical homogeneous sections the coordinate $x\in[0,\pi)$ has the interpretation of the latitude and all familiar identifications of points, as in the case of spherical coordinates are asuumed. In what follows we shall denote $x$ by $\theta$.

 In spherical case we have $H=2\al\sin^2(\theta/2)$ and  $D=\sin(\theta)$. With the ansatz
 \[
 \Psi(t,\theta,\vp,x_3)=e^{-i\w t}e^{ik_3x_3}\cdot \psi(\theta,\vp),
 \]
we find
\begin{equation}
\Box\Psi=\left[ C_2-(1+\al^2)\w^2+k_3^2\right] \Psi=0,
\end{equation}
where $C_2=(L_1)^2+(L_2)^2+(L_3)^2$ is the Casimir operator of a modification of the algebra the SO(3) of (symmetry) generators:
\begin{eqnarray}
L_+&=z^2 \D +  \Db +a z ,\\
L_-&= -\D - \zb^2 \Db +a \zb,\\
L_3&=z\D-\zb\Db+a,
\end{eqnarray}
with $L_\pm=L_1\pm iL_2$, and
\[
z=\tan{(\theta/2)}e^{i\vp}.
 \]
The above operators fulfill the SO(3) relations:
\[
[L_+,L_-]=2L_3,\qquad [L_\pm,L_3]=\mp L_\pm.
  \]
The task is now to determine the functions, $\psi(\theta,\vp)$,
which in order to make the d'Alembert equation separable must be eigenfunctions of the Casimir operator $C_2$. This operator (as well as $L_1, L_2, L_3$) is a selfadjoint operator on the Hilbert space of square-integrable functions with respect to the product
\[
(f,g)=\int \sin\theta\, d\theta d\vp\, \ba{f}\, g.
\]
Let us investigate the sectors consisting of eigenvectors of  $C_2$ to the eigenvalue $\la(\la+1)$ and of $L_3$ to eigenvalues in $[-\la,\la]$, where $\la$ is a positive integer or half-integer\footnote{Only the integer or half integer values of $\la$ are allowed, for otherwise one would be lead to standard contradictions (eg. states of negative norm, after sufficiently many applications of $L_+$).
}.
The lowest states in each of the sector, $\psi_{-\la}$, which are annihilated by $L_-$ with the eigenvalue $-\la$ of $L_3$, and the highest states $\psi_\la$ (annihilated by $L_+$ with the eigenvalue $\la$ of $L_3$) can now be determined easily:
\begin{eqnarray}
\psi_{-\la}&=(1+z\zb)^{-\la} \zb^{\la+a},\\
\psi_{\la}&=(1+z\zb)^{-\la} z^{\la-a}.
\end{eqnarray}
These function must be periodic in $\vp$, with identical values for $\vp=0$ and $\vp=2\pi$, and therefore $a+\la\equiv m$ must also be an integer. Now it is easy to see, that either there holds
\begin{eqnarray}
m&\geq 0,\\
m&\leq 2\la,
\end{eqnarray}
or the function $\psi_{-\la}$ will not be square integrable on the sphere (in $x,\vp$) because of singularities at $\theta=0$ or $\theta=\pi$ respectively. Consequently, for a given value of $\la\in\tfrac{\mathbb N_0}{2}$ there remains a freedom to choose $m\in[0,2\la]$. The parameter $a$, and therefore also the frequency $\w=-a/\al$, is fixed by $\la$ and $m$:
\[
a=-\la+m\in[-\la,\la].
\]
The frequencies are discrete, and their spacing is equal (as for the harmonic oscillator):
\[
\w=\frac{m-\la}{\al}.
\]
Such a quantization of frequencies has an alternative, kinematical (gauge) derivation pointed out long time ago by Mazur \cite{Ma,Ma_c}.

Within each sector of fixed $\la$ and $m$, there is a family of $2\la+1$ functions ${_m}S^\la_k(\theta,\vp)$, which are eigenfunctions of $L_3$ to the eigenvalue $k$, with $\tfrac{\mathbb Z}{2}\ni k\in[-\la,\la]$. These functions are generated by a successive application of $L_+$ to
\[
\psi_{-\la}={_m}S^\la_{-\la}=(1+z\zb)^{-\la}\zb^m.
\]
 The only role of the wave equation is to determine the momentum in $x_3$ direction:
\begin{equation}
k^2_3=(1+1/\al^2)(-\la+m)^2-\la(\la+1).
\end{equation}
Evidently, as a consequence of the discreteness of $\la$ and $m$ the momenta, $k_3$, become discrete too. For each $m$ positivity of $k_3^2$ excludes a certain interval of $\la$'s (see Fig. 2).

\begin{figure}[h]
\includegraphics[scale=0.6]{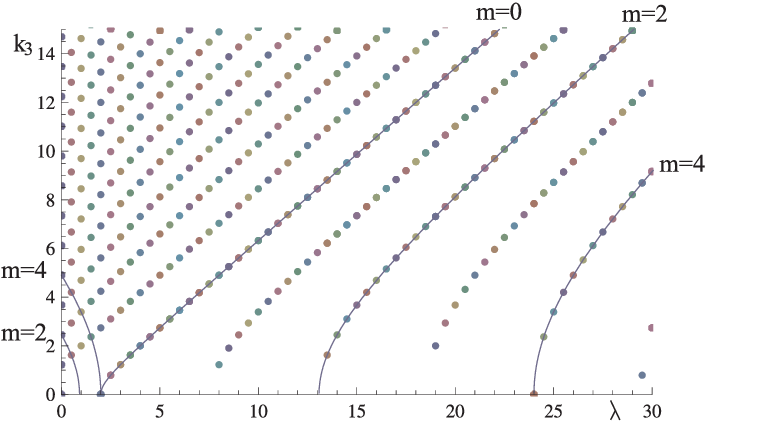}
\caption{Momenta in $x_3$ direction in spherical case as a function of $\la$ for $m=0,2,4$.}
\label{firstfig}
\end{figure}

We finally note, that the functions we have found, ${_m}S^\la_k$ are also known as the monopole harmonics, ${_a}Y^\la_k$, or the spin-$a$ spherical harmonics \cite{Tamm,Dray}, for there holds
\[
{_{m-\la}}Y^\la_{k}={_m}S^\la_k.
\]
The eigenspace $\la=\frac{1}{2}$, $m=0$ consists of two functions first encountered by Dirac \cite{Dirac}  in the context of magnetic monopoles,
\begin{equation}
{_0}S^{{1/2}}_{-{1/2}}=\cos(\theta/2), \qquad {_0}S^{{1/2}}_{{1/2}}=\sin(\theta/2) e^{i\vp}.
\end{equation}

\subsection*{Summary of spherical case}
The solutions of the d'Alembert equation have the form:
\[
\Psi(t,x,\vp,x_3)=e^{-i\w t}e^{ik_3x_3}\cdot (_mS^\la_k)(\theta,\vp),
\]
with the following restrictions/specifications on the parameters of the function $(_mS^\la_k)$:
\begin{table}[h]
\caption{Summary of the spherical case.}\label{eqtable}
\renewcommand\arraystretch{1.5}
\noindent\[
\begin{array}{|l|c|}
\hline
&spherical \ case\\
\hline
sector, {\la}&\la \in \frac{\mathbb N_0}{2} \\
\hline
eigenvalue \, of\, J_0, \, {k}&\frac{\mathbb N_0}{2} \ni k\in[-\la,\la]\\
\hline
eigenvalue \, of \ C_2&\la(\la+1)\\
\hline
extremal \, vector, \, (_m S^\la_{-\la}) & (1+z\zb)^{-\la} \zb^m\\
\hline
& z=\tan(\theta/2)e^{i\vp}
\\
\hline
vectors\ {_mS^\la_k} \ generated \ by & L_+ \\
\hline
parameter\ m& \mathbb N_0\ni  m\in[0,2\la]\\
\hline
parameter \ {a}& a=m-\la\\
\hline
frequency \ \w& \w=-\frac{m-\la}{\al}\\
\hline
\end{array}
\]
\end{table}

The values of $\la$ and $m$ specify completely the frequency and, by the wave equation, the momentum in the $x_3$-direction:
\begin{equation}
k_3=\pm\sqrt{(1+1/\al^2)(\la-m)^2-\la(\la+1)}. \label{allowed_k}
\end{equation}
For each $m$ positivity of the expression under the square root excludes of interval of $\la$'s.

\section{Remarks and outlook}
In this chapter we have constructed full families of solutions of the d'Alembert equation on G\"odel-type spacetimes with Lobachevsky (hyperbolic) and spherical homogeneous sections\footnote{An algebraic solution can also be given in the case of flat homogeneous sections. We refer the interested reader to \cite{DFS}.}. The families are full in the sense, that any additional solutions would violate the imposed regularity conditions at $x=0$ ($\theta=0$) or at $x=\infty$ ($\theta=\pi$) in Lobachevsky (spherical) case. For example, in the spherical case it is necessary for the solutions be $2\pi$-periodic in $\vp$. Consequently, with our ansatz \eqref{separation}, the wave equation reduces to a second order ODE in $\theta$. Solutions of this ODE, not present in our family,  diverge as $(\sin \theta/2)^{-\mu}$ at $\theta=0$ (or as $(\cos \theta/2)^{-\mu}$ at $\theta=\pi$), with $\mu\geq 1$.

With the families being full in the above sense  it is important to note that this does not imply completeness, understood as a possibility to express arbitrary ``initial data'' $\{f,\partial_t f\}$ compactly supported on the hypersurface $t=t_0$ (say, in a region, where this surface is still spacelike) isometrically via a sum of the solutions from our family, restricted to $t=t_0$.
 
 Indeed we have seen in the spherical case, that the momenta in $x_3$ direction are discrete (although there are no a priori boundaries/periodicities in this direction). We regard it to be an indication of a possible incompleteness of the family of solutions. Such a behavior appears typical for spacetimes with closed timelike curves. Consider, for instance the simplest example of a 2-dimensional flat spacetime  $ds^2=dt^2-dx^2$ with periodic time $t$ ($t=0$ identified with $t=T$) and $x$ unconstrained ($x\in\mathbb R$). The solutions of the wave equation in this spacetime, as a consequence of this wave equation, are necessarily periodic in $x$ with the period $T$. If we imposed periodicity of the spacetime in $x$, e.g. by identifying $x=0$ with $x=L$, then - depending on the value of $L/T$ the wave equation would have no (non-constant) solutions  for irrational $L/T$, or would possess a complete set of solutions (in the sense considered above) for rational $L/T$'s.

  Whether the same behavior occurs in G\"odel-type spacetimes is yet unclear (see e.g. Eq. \eqref{allowed_k} for the allowed values of $k$). The important question of the completeness of the families of solutions of the wave equation in G\"odel-type spacetime does not seem to have been investigated in the literature.

Finally, let us express a hope that a precise determination of the properties of waves considered here might find an application in the so-called analogue gravity models, where phase perturbations of the order parameter in given configurations of the Bose-Einstein condensates or small perturbations of given hydrodynamic flows behave as scalar massless waves in curved spacetimes (see \cite{stone} for an example involving vorticity of the background flow).

\bibliographystyle{amsalpha}

\begin{appendix}
\section{Preliminaries on the  Lobachevsky space}\label{prelims}
The 2-dimensional Lobachevsky space (known also as the hyperbolic plane, $\mathbb H_2$) is the space defined as the hypersurface $(x_0)^2-(x_1)^2-(x_2)^2=1$ of the 1+2 dimensional Minkowski space (with coordinates $x_0,x_1,x_2$). The intrinsic metric of this space is given by
\[
ds^2=dx^2+\sinh^2(x)\, d\vp^2,
\]
with $x\in[0,\infty)$, $\vp\in[0,2\pi)$. (This map covers the whole space.) The three Killing vector fields,
\begin{align}
K_1&=\sin(\vp)\D_x+\coth(x) \cos(\vp) \D_\vp\\
K_2&=-\cos(\vp)\D_x+\coth(x) \sin(\vp) \D_\vp\\
K_0&=\D_\vp,
\end{align}
by analogy to rotations on a sphere, when multiplied by $-i$ lead to the generators of the corresponding symmetries, which are differential operators on the Hilbert space of square integrable functions on $\hh$, with the standard product
 \begin{equation}\label{product}
(f,g)=\int \overline{f}g\,
\sinh(x)\, dx\, d\vp.
\end{equation}
These generators, denoted here by $L_i$, fulfill the commutation relations of the algebra of SU(1,1) generators:
\begin{align}
[ L_0,L_1]&=iL_2\\ \,
[ L_2,L_0] &=iL_1\\ \,
[ L_1,L_2] &=-iL_0,
\end{align}
and commute with the quadratic Casimir operator
\begin{equation}
C_2=L_0^2-L_1^2-L_2^2.
\end{equation}
The group SU(1,1) is not compact, and therefore the representation theory is richer than that of the rotation group.  Let us finally note,  that
\begin{equation}
C_2=\D_x^2 +\coth(x)\D_x+\frac{1}{\sinh^2(x)}
\D^2_\vp=\frac{1}{\sqrt h}\D_i\left( h^{ij} \sqrt{h} \D_j\right)
\end{equation}
is just the Laplace-Beltrami operator on the Lobachevsky space and therefore \emph{here $C_2\leq 0$}.

\end{appendix}

\end{document}